 \documentclass[11pt,twocolumn]{article}

 \pdfoutput=1

 \usepackage[english]{babel}
\usepackage{graphicx} % for .epsi formats
\usepackage{amsmath}

\usepackage{enumitem}

\usepackage{float}
\usepackage{ifthen}
\usepackage{hyperref}

\usepackage{titlesec}

\titlespacing*{\section}{0pt}{0.2\baselineskip}{\baselineskip}

\begin{document}

%-------------------------------------------------------------------

\pagestyle{empty}

%\onecolumn

\twocolumn[
%\title
\begin{center}
{\bf \huge 
Definition of Total Energy budget equation in terms of moist-air Enthalpy surface flux.
}\\
\vspace*{3mm}
%\author
{\Large \bf by Pascal Marquet } {\Large (WGNE Blue-Book 2015)}. \\
%{\huge by Pascal Marquet}. \\
\vspace*{2mm}
{M\'et\'eo-France. CNRM/GMAP.
 Toulouse. France.}
{\it E-mail: pascal.marquet@meteo.fr} \\
\vspace*{1mm}
\end{center}
]

%\corraddr{Pascal MARQUET, DPr\'evi/Labo, M\'et\'eo-France, 42 av. G. Coriolis, 31057 Toulouse CEDEX 01, France.\\ Web site: http://perso.numericable.fr/$\sim$pmarquet/  ; E-mail: pascal.marquet@meteo.fr }

%\date{\today}
%\date{March, 2015}

%\maketitle

%-------------------------------------------------------------------

% -----------------------
 \section{\underline{\Large Motivations.}} % (Section 1)
% -----------------------
\vspace{-4mm}

The way moist-air surface heat flux should be computed in atmospheric science is still a subject of debate.
It is explained in Montgomery (1948, M48), Businger (1982, B82), more recently Ambaum (2010, \S 3.7), that  uncertainty exists concerning the proper formulation of surface heat fluxes, namely the sum of ``sensible'' and ``latent''  heat fluxes, and in fact concerning these two fluxes if they are considered as separate fluxes.

It is shown in M48 that eddy flux of moist-air energy must be defined as the eddy transfer of moist-air specific enthalpy $h = e_{\rm int} + p/\rho = e_{\rm int} + R \; T$, where $e_{\rm int}$ is the internal energy.
However, the way the moist-air specific enthalpy $h$ is computed in M48 still depends on some arbitrary assumptions concerning reference value of dry-air or water-vapour enthalpies, which are set in M48 to arbitrary conventional values at a finite reference temperatures different from $0$~K.

Consequences of these arbitrary assumptions are studied at length in B82, though without succeeding in computing the moist-air enthalpy in an absolute way.

Issues addressed in M48 and B82 can be overcome by using the specific thermal enthalpies derived in Marquet (2015, M15) for N$_2$, O$_2$ and H$_2$O, namely for the main components of moist air.

In this article this approach is taken to show that Third-law based values of moist-air enthalpy fluxes is the sum of two terms.
These two terms are similar to what is called ``sensible'' and ``latent'' heat fluxes in existing surface energy budget equation, but a new kind of ``latent heat'' is emerging in the definition of the moist-air enthalpy flux.
Some impacts of this new ``latent heat'' flux are described in this brief version of a paper to be submitted to the QJRMS.
\vspace{-1mm}

% -----------------------
 \section{\underline{\Large The energy budget equation.}} % (Section 2)
% -----------------------
\vspace{-3mm}

Only three kinds of specific energies can be defined in atmosphere if nuclear or chemical reactions are not considered: i) the kinetic energy $e_k = (u^2+v^2+w^2)/2\:$; ii) the potential energy $\phi = g \: z\:$; iii) the moist-air thermal internal energy $e_{\rm int} = h -  p / \rho$ or enthalpy $h = e_{\rm int} + p / \rho$ which are both associated with the First-Law of Thermodynamics.

The total energy equation of a unit mass of moist air is computed for the sum $e_{\rm tot} = e_{\rm k} +  \phi + e_{\rm int}$ by adding the three local equations for $e_{\rm k}$, $\phi$ and $e_{\rm int}$, yielding
\vspace*{-3mm}
\begin{align}
 \frac{\partial}{\partial t}
 \left[ \,
     \rho
        \left( \,
          e_{\rm int} + \phi + e_k
        \, \right)
 \, \right]
  & = - 
  \boldsymbol{\nabla} .
 \left[ \,
     \rho
        \left( \,
          h + \phi + e_k
        \, \right)
  {\bf U}
 \, \right]
  + \rho \: \dot{q} \, ,
\nonumber
\end{align}

\vspace*{-4mm}
\noindent where $\dot{q}$ is a notation for impacts of radiation and other local sinks and sources of energy.
The specific enthalpy $h$ appears in the divergence term because the sum of 
$- \: {\bf U} . \boldsymbol{\nabla}(p)$ and 
$-p \: \boldsymbol{\nabla} . {\bf U}$ 
in equations for $e_{\rm k}$ and $e_{\rm int}$, respectively,
is equal to 
$-\boldsymbol{\nabla} . \: ( \: p \: {\bf U}) =
 -\boldsymbol{\nabla} . \: [ \: \rho  \: (p/\rho) \: {\bf U} ]$,
 thus leading to the local definition $e_{\rm int} + p / \rho \equiv h$.

%===============
% Figure (1) :
%===============
%=====================
% Figure  : Mean/Eddy
%===============================================================
% Voir la procedure sous LXGMAP4 : 
% .../DOCUMENTS_SAVE/Norme_Am/Mahfouf_Bilodeau/zprog_poids.sh
%===============================================================
\begin{figure}[hbt]
\centering
\includegraphics[width=0.5\linewidth]{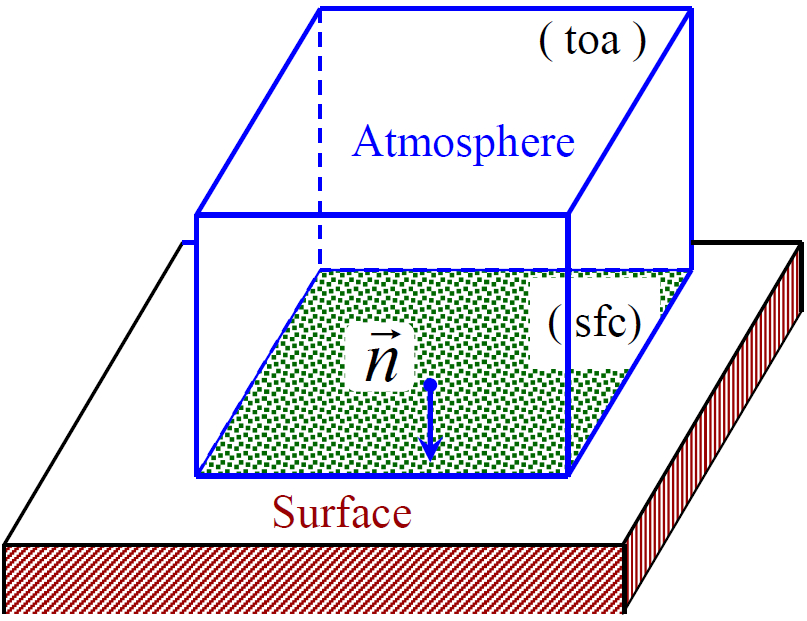}
\vspace{-3mm}
\caption{\small \it 
A column of atmosphere above Earth's surface.
\label{fig_1}}
\end{figure}
The total energy budget equation is then computed by integrating this local equation over an infinite vertical column of atmosphere (see Fig.\ref{fig_1}), leading to
\vspace*{-1.5mm}
\begin{equation}
 \frac{\partial E_{\rm \, tot}}{\partial \, t} 
   \: = \:
  \left\langle \:  \rho
        \left( \,
          h + \phi + e_k
        \, \right) w \:  
  \right\rangle
  \; + \: \dot{Q}
  \: ,
 \nonumber
\end{equation}

\vspace*{-4mm}
\noindent where lateral fluxes have been neglected.
The term $\left\langle ( ... ) \right\rangle$ denotes the surface average of
$\rho \: ( h + \phi + e_k ) \: w$,
and the integral of ``$ \, \rho \: \dot{q} \:$'' is written as $\dot{Q}$. 

The exchange of total energy between the column and the surface thus depends on 
$\left\langle \: \rho \: ( h + \phi ) \: w \: \right\rangle$
which can be rewritten by using Reynolds average 
$\overline{(...)}$ and perturbation $(...)'$ terms, 
leading to the local turbulent fluxes of specific enthalpy and potential energy
\vspace*{-1.5mm}
\begin{align}
  F_h
  & \: \equiv \; 
   \overline{ (\rho \: w)' \: h' }
   \; \: \approx \; 
   \overline{\rho\,} \: \overline{\, w' \: h' \,}
  \: ,
 \label{eq_Budget_F_h1} \\
   F_{\phi}
  & \: \equiv \; 
   \overline{ (\rho \: w)'  \: \phi' }
   \; \: \approx \; 
   \overline{\rho\,} \: \overline{\, w' \: \phi' \,}
  \: .
 \label{eq_Budget_F_phi}
\end{align}

\vspace*{-3mm}
\noindent Conclusions are the same as in M48 or B42: from (\ref{eq_Budget_F_h1}), it is needed to know specific values of moist-air thermal enthalpy $h$, in order to compute $\overline{\, w' \: h' \,}$.

% -----------------------
 \section{\underline{\Large The specific thermal enthalpy.}} % (Section 3)
% -----------------------
\vspace{-3mm}

The moist-air enthalpy is equal to the weighted average of individual (perfect gas) values for dry air, water vapour, liquid water and ice species, leading to $ h  =  q_d \: h_d  + q_v \: h_v  + q_l \: h_l  + q_i \: h_i$. 
This sum can be computed according to M15 (with different algebra), leading to
\vspace*{-3mm}
\begin{align}
 \!\!
 h \; = & \; h_{\rm ref} 
  \: + \:  c_{pd} \; T 
  \: + \:  L_h \: q_t
  \: - \:  L_{\rm \, vap} \; q_l
  \: - \:  L_{\rm \, sub} \; q_i
  \: ,
 \label{eq_h} \\
 & \mbox{where} \;\;
  L_{\,\rm sub} \: (T) \; \equiv \; h_v(T) \: - \: h_i(T)
   \: ,
  \nonumber\\
 & \hspace{12mm}
   L_h \: (T)  \; \equiv \; h_v(T) \: - \: h_d(T)
   \: ,
 \nonumber \\
 & \hspace{12mm}
   L_{\,\rm vap} (T) \; \equiv \; h_v(T) \: - \: h_l(T)
  \: ,
 \nonumber
\end{align}

\vspace*{-4mm}
\noindent 
where $q_t = q_v + q_l + q_i$ is the total water content.
The atent heats are in fact ``differences in enthalpies''. 
They only depends on temperature and on some reference values, with for instance 
$L_h (T) \;  = \; L_h(T_r) \: + \: (c_{pv} - c_{pd} ) \:  ( \, T - T_r \, )$ and 
$L_h(T_r) = (h_v)_r \: - \: (h_d)_r$.

Issues reported in M48 an B82 can thus be understood by a need to know $(h_d)_r$ and $ (h_v)_r$ in order to compute $L_h(T_r)$, then $L_h(T)$, and finally ``$h$'' via (\ref{eq_h}).
Dry-air and water-vapour reference thermal enthalpies are computed in M15 at $0$~C: 
$(h_d)_r  \approx 530 \mbox{~kJ} \mbox{~kg}^{-1}$ and  $(h_v)_r  \approx  3133  \mbox{~kJ} \mbox{~kg}^{-1}$.
The moist-air reference enthalpy is thus equal to $h_{\rm ref} =  (h_d)_r - c_{pd} \: T_r \approx 256$~kJ~kg${}^{-1}$.
It is a true constant (whatever $T_r$ may be in atmospheric range of $T$ with constant $c_{pd}$) and it does not impact on gradient or flux computations.

%===============
% Figure (2) :
%===============
%================================================
% Figure : L_h(T) versus L_vap(T) and L_sub(T)
%===============================================================
% Voir la procedure sous LXGMAP4 : 
% .../DOCUMENTS_SAVE/Surf_budget/zproc_cp_rot_vibr2.sh
%===============================================================
\begin{figure}[hbt]
\centering
\includegraphics[width=0.99\linewidth]{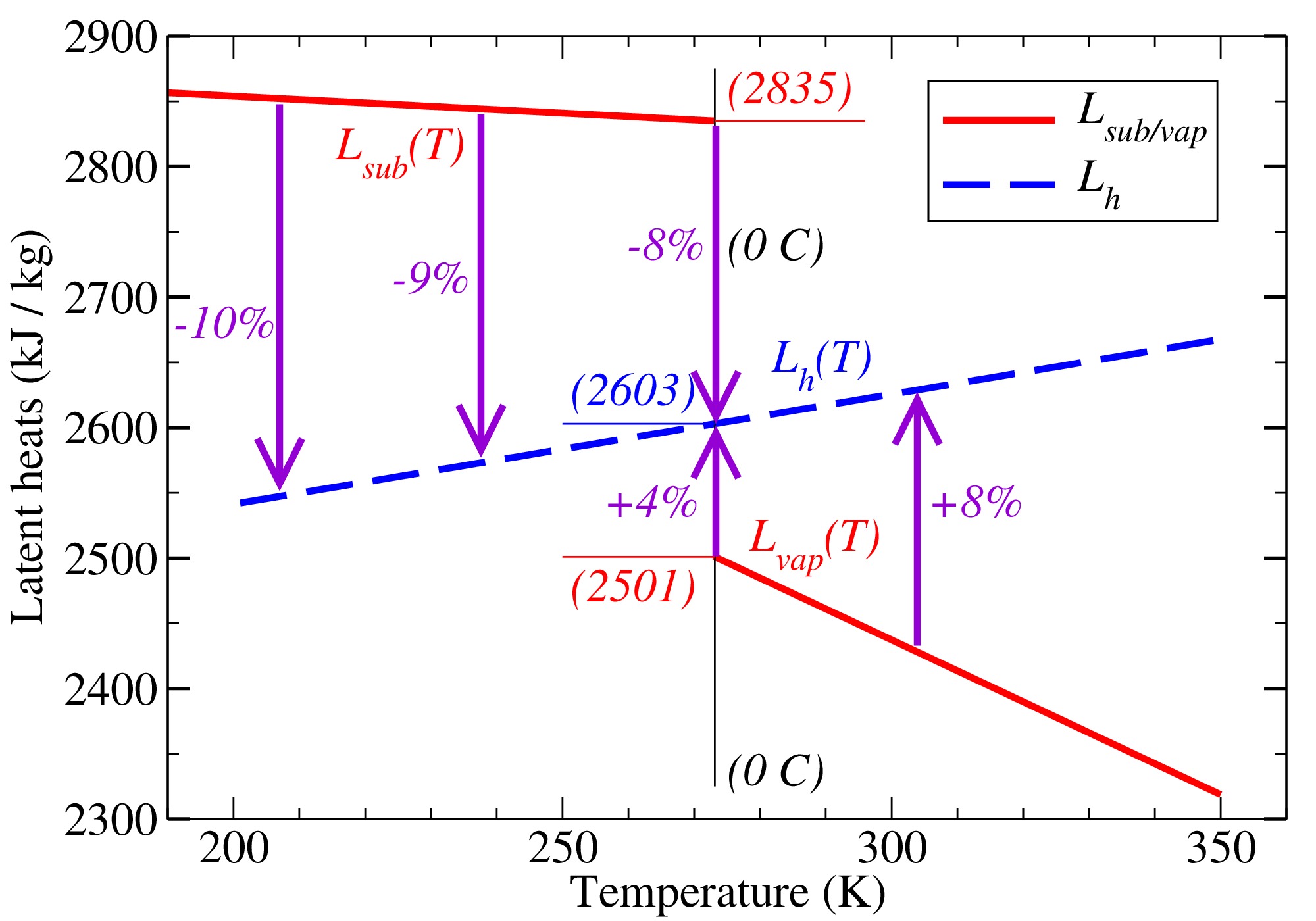}
\vspace{-8mm}
\caption{\small \it 
Comparison of $L_h(T)$ with $L_{\,\rm vap}(T)$ and $L_{\,\rm sub}(T)$.
Unit are in kJ~kg${}^{-1}$ for latent heats, in K for $T$.
\label{fig_2}}
\end{figure}

\vspace*{-3mm}
\noindent 
Changes of $L_{\,\rm sub}$, $L_h$ and $L_{\,\rm vap}$ with absolute temperature are compared in Fig.\ref{fig_2}. 
The dashed straight line represents $L_h(T)$.
It  is continuous at $0$~C and is in between solid lines representing $L_{\,\rm sub}(T)$ and $L_{\,\rm vap}(T)$.

% -----------------------
 \section{\underline{\Large The moist-air enthalpy flux.}} % (Section 4)
% -----------------------
\vspace{-4mm}

The moist-air thermal enthalpy fluxes $F_h \approx \overline{\rho\,} \: \overline{\, w' \: h' \,}$ can be computed with $h$ defined by (\ref{eq_h}), yielding
\vspace*{-3mm}
\begin{align}
  F_h
  & \: = \; 
   c_p \: F_T
   \; + \; 
   L_h \: F_v
  \nonumber \\
   & \quad 
   \; + \; 
   \left( L_h - L_{\,\rm vap} \right) \: F_l
   \; - \; 
   \left(  L_{\,\rm sub} -  L_h  \right) \: F_i
  \: ,
 \label{eq_Budget_F_h}
\end{align}

\vspace*{-4mm}
\noindent 
where moist value of $c_p$ is considered and where
$F_T \approx \overline{\rho\,} \: \overline{\, w' \: T' \,}$,
$F_v \approx \overline{\rho\,} \: \overline{\, w' \: q'_v \,}$ ,
$F_l \approx \overline{\rho\,} \: \overline{\, w' \: q'_l \,}$ and
$F_i \approx \overline{\rho\,} \: \overline{\, w' \: q'_i \,}$
are turbulent fluxes of temperature and water species.

Clearly, $L_h - L_{\,\rm vap}$ and $L_{\,\rm sub} -  L_h$ represent about $4$ to $8$~\% of $L_{\,\rm vap}$, and about $-8$ to $-10$~\% of $L_{\,\rm sub}$.
The second line of (\ref{eq_Budget_F_h}) can thus be neglected  with an accuracy better than $10$~\% and $F_h \approx c_p \: F_T + L_h \: F_v$.
However, if liquid water of ice contents exist, it is easy to avoid any approximation by taking into account the second line and the small terms depending on $F_l$ or $F_i$.
\vspace{-5mm}

% -----------------------
 \section{\underline{\Large Numerical evaluations.}} % (Section 5)
% -----------------------
\vspace{-4mm}

From  Fig.\ref{fig_2} and for Earth's surface temperatures from $0$ to $30$~C, $L_h$ is about 6~\% larger than $L_{\,\rm vap}$ in average.
The change of $L_{\,\rm vap}(T)$ by $L_h(T)$ should thus lead to an increase of about 6~\% for existing surface latent heat fluxes which are about $100$~W~m${}^{-2}$ in average.
Impact of using  $L_h(T)$  on global energy budget of atmosphere should thus be of the order of $+6$~W~m${}^{-2}$ in average.
But impact larger than $+50$~W~m${}^{-2}$ could exist locally.

These values have been confirmed by using short-range forecasts of ARPEGE NWP model, and in particular for convective or frontal regions.
\vspace{-1mm}

% -----------------------
 \section{\underline{\Large Conclusions.}} % (Section 4)
% -----------------------
\vspace{-4mm}

It has been shown that computations of budget of total energy imply computations of turbulent fluxes of specific enthalpy, and that it is possible to achieve the program started in M48 and B82 by computing these turbulent fluxes of enthalpy (\ref{eq_Budget_F_h}) by using reference values of enthalpies derived in M15.

The surface enthalpy flux (\ref{eq_Budget_F_h}) may replace what are commonly called ``latent'' and ``sensible'' heat fluxes, with the use of a new ``latent heat'' $L_h(T)$ which is the difference in enthalpies of dry air and water vapour.
Impacts of using $L_h(T)$ is of the order of $+6$~W~m${}^{-2}$ in average, and more than $+50$~W~m${}^{-2}$  locally.

$L_h(T)$ represents physical processes like evaporations over oceans where water vapour enters atmospheric parcels of moist air, this meaning a decrease in $q_d$ at the expense of an increase of $q_v$, in other words a replacement of dry air by water vapour.

A striking feature imposed by the definition of the moist-air enthalpy (\ref{eq_h}) is that $L_h(T)$ is continuous at $0$~C.
This is in direct contradiction to the usual definitions of latent heat fluxes, assumed to be equal to $L_{\,\rm vap} \: F_v$ over liquid water and $L_{\,\rm sub} \: F_v$ over ice.

\vspace{1mm}
\noindent{\large\bf \underline {References}}
\vspace{0mm}

\noindent{$\bullet$ M.~H.~P. Ambaum} {(2010)}.
{\it Thermal physics of the atmosphere.}
Advancing weather and climate science.
Wiley-Blackwell.
John Wiley and sons. Chichester.

\noindent{$\bullet$ J.~A. Businger} {(1982)}.
 The fluxes of specific enthalpy, sensible 
 heat and latent heat near the Earth's surface.
{\it J. Atmos. Sci.}
{\bf 39} (8):
p.1889--1892.

\noindent{$\bullet$ P. Marquet} {(2015)}.
 On the computation of moist-air specific 
 thermal enthalpy.
{\it Q. J. R. Meteorol. Soc.}
{\bf 141} (686):
p.67--84.

\noindent{$\bullet$ R.~B. Montgomery} {(1948)}. 
 Vertical eddy flux of heat in the atmosphere.
{\it J. Meteorol.}
{\bf 5}, (6): 
p.265--274.

%-------------------------------------------------------------------

\end{document}